\documentclass[12pt]{article}
\pdfoutput=1
\usepackage[a4paper]{geometry}
\usepackage{amsmath,amssymb,amsfonts,latexsym}
\usepackage{graphicx}  
\usepackage[colorlinks=true, pdfstartview=FitV, linkcolor=blue, citecolor=blue, urlcolor=blue]{hyperref}

\newcommand{\bra}{\langle} 
\newcommand{\ket}{\rangle}
\newcommand{\rstr}{\bigg\vert}
\newcommand{\ep}{\qquad {\vrule height 10pt width 8pt depth 0pt}}
\newcommand {\mD}{{\bf D}}
\newcommand {\mO}{{\bf O}}
\newcommand {\fg}{{\mathfrak g}}
\newcommand {\bR}{{\mathbb R}}
\newcommand {\bN}{{\mathbb N}}

\newcommand {\bZ}{{\mathbb Z}}
\newcommand {\bI}{{\mathbb I}}

\newcommand {\bC}{{\mathbb C}}

\newcommand {\bH}{{\mathbb H}}

\newcommand {\cA}{{\cal A}}

\newcommand {\cC}{{\cal C}}

\newcommand {\cG}{{\cal G}}

\newcommand {\cO}{{\cal O}}
\newcommand {\cS}{{\cal S}}

\newcommand {\cU}{{\cal U}}
\newtheorem{theorem}{Theorem} [section]
\newtheorem{lemma}[theorem]{Lemma}
\newtheorem{propo}[theorem]{Proposition}
\newtheorem{defi}[theorem]{Definition}
\newtheorem{corollary}[theorem]{Corollary}

\begin{document}
\title{A constant of quantum motion in two dimensions in crossed magnetic and electric fields\\
{\normalsize \it We dedicate this work to the memory of Pierre Duclos} \bigskip }
\author{Joachim Asch\thanks{CPT-CNRS UMR 6207, Universit\'e du Sud, ToulonÐVar, BP 20132,F--83957 La Garde Cedex, France, e-mail:asch@cpt.univ-mrs.fr}  , C\'edric Meresse\thanks{CPT-CNRS UMR 6207, CNRS Luminy, Case 907 13288 Marseille Cedex 9, France, e-mail : meresse@cpt.univ-mrs.fr}}
\date{06.09.2010}
\maketitle

\abstract{We consider the quantum dynamics of a single particle in the plane under the influence of a constant perpendicular magnetic and a crossed electric potential field. For a class of smooth and small potentials we construct a non-trivial invariant of motion. Do to so we proof that the Hamiltonian is unitarily equivalent to an effective Hamiltonian which commutes with the observable of kinetic energy.
}

\section{Introduction}
Consider a particle of mass $m$ and charge $e$ in the plane under the influence of a constant magnetic field of strength $B$ and an electric potential. We choose  the units of magnetic length $\sqrt{\frac{\hbar}{\vert e B\vert}}$, the gyration time $\frac{m}{\vert e B\vert}$, and the energy gap $\frac{\hbar\vert e B\vert}{m}$. The dynamics are generated by 

\[H=H_{La}+ V\quad{\rm in } \quad L^{2}(\bR^{2})\]
with
\[H_{La}=\frac{1}{2}\left(-i\nabla-\frac{q^{\perp}}{2}\right)^{2}.\]
with  operator core the Schwartz space $\cS(\bR^{2})$.  $V$ is the multiplication operator by a function $V(q)$ and $\left(q_{1}, q_{2}\right)^{\perp}:=(-q_{2},q_{1})$. 
 For the gaussian
\[g(q):=e^{-\frac{q^{2}}{2}}\qquad (q\in\bR^{2})\]
we consider the class of functions defined by convolution with a real valued finite measure $\mu$, $g\ast\mu(q):=\int_{\bR^{2}}g(q-q^{\prime})d\mu(q^{\prime})$ :

\[{\bf\cG}:=\{V:\bR^{2}\to\bR; V=g\ast\mu,\int_{\bR^{2}}d\vert\mu\vert<\infty\}.\]

Our main result is that there exists a non-trivial integral of motion;  thus, in this weak sense, the two-dimensional system is integrable:

\begin{theorem}\label{thm:main} For $V\in\cG$  small enough there exists a unitary operator $U$ such that  
\[\left\lbrack U^{-1}\left(H_{La}+V\right)U, H_{La}\right\rbrack=0.\]
In particular:  $U H_{La}U^{-1}$ is an invariant of the flow $e^{-i H t}$ for all $t\in\bR$.
\end{theorem}

\begin{figure}[hbt]
\centerline {
\includegraphics[width=7cm]{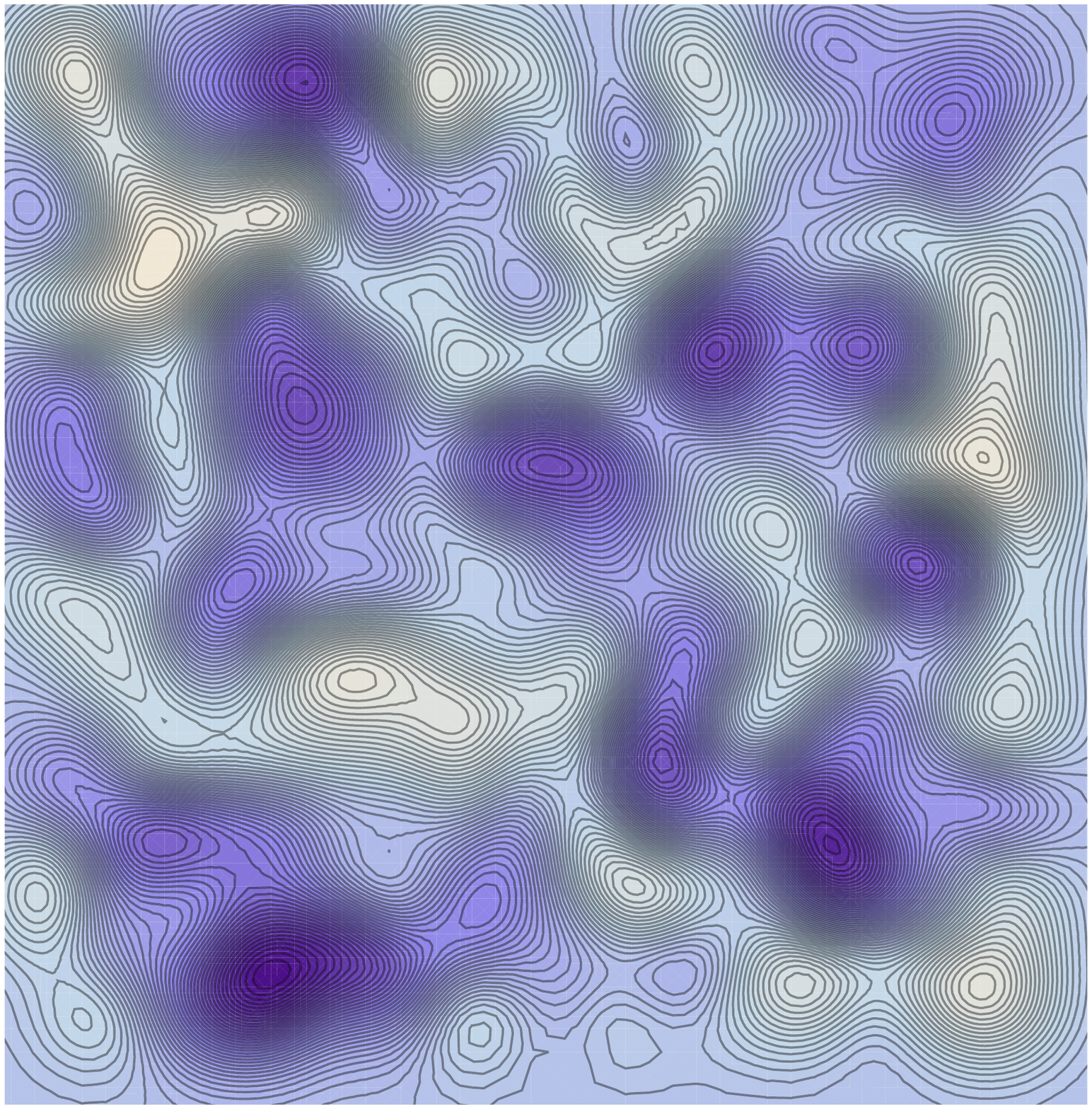}
}
\caption{$V(q)=\sum_{i\in\lbrack-10,10\rbrack^{2}}u(i)g(q-i)$,  u(i) i.i.d random variables}
\label{fig:potential}
\end{figure}

The meaning of ``small enough'' will be made precise in the sequel. A potential in $\cG$ is depicted in figure \ref{fig:potential}.

It is folklore in plasma physics that in slowly varying fields the classical particle gyrates on a cycloid whose center drifts along the contour lines of the averaged potential and whose kinetic energy is an approximate invariant up to a certain time \cite{cb,nei}. On the other hand classically  chaotic motion may occur in relevant regimes of parameters \cite{gwno}.

In the realm of quantum physics the corresponding invariance is, in the large magnetic field limit, an essential ingredient for the current understanding of the integer quantum Hall effect \cite{l,bsbe, ass, hs,  gks, cc, abj} while the interesting physics happen in the lowest Landau level. Coupling between Landau bands may, however,  lead to non negligible effects \cite{pg}. 

Several methods to construct an approximatively  invariant subspace, making precise the notion of lowest Landau level, and an effective Hamiltonian are known, see \cite{n,t,bdp}. They may lead to estimates valid to exponential order in a small parameter and valid for exponentially long times. See \cite{bg} for an application of these ideas to a propagation problem. Our aim here is to study all Landau levels simultaneously as well as the possibility to go to the limit of all times. 

Our strategy is to employ a superconvergent iterative partial  diagonalization procedure  which was originally introduced in quantum problems to discuss stability of non-resonant time periodic problem \cite{b,c,ds}; see \cite{ade} for an application  to a condensed matter problem. It coincides in first order with the above mentioned ``space adiabatic'' algorithms. In particular  the effective Hamiltonian restricted to the lowest Landau level is in first order
$\bra V\ket(x, D)$, a pseudodifferential operator whose symbol is the potential averaged over the Landau orbits. At  higher orders the algorithm differs; on one hand it exhibits quadratic convergence, on the other hand it is an unsolved, and to our opinion important problem,  whether our higher order effective operators are pseudodifferential.

 The partial diagonalization procedure  is roughly described as follows:  the diagonal part of an operator $H$ with respect to a fixed,  orthogonal, mutually disjoint family of projections $\{ P_{n}\}$ is defined by $\mD H:=\sum_{n}P_{n}H P_{n}$. In fact, as the dimension of $P_{n}$ is not supposed to be one, or even to be finite, $\mD H$ is block-diagonal. Suppose that the off-diagonal part $\mO H:=H-\mD H$ is small with respect to $\mD H$. The equation
\[e^{W}H e^{-W}-\mD H=\cO\left(\Vert \mO H\Vert^{2}\right)\]
is satisfied by an antiselfadjoint operator $W$ of order $\cO\left(\Vert \mO H\Vert\right)$ solving
\[\mO H+\lbrack W, \mD H\rbrack=0.\]
If one takes $P_{n}$ the projections on the Landau levels then,  because of the gap, a solution of this equation can be found  if the coupling between the bands decays sufficiently fast. This is the case for the potentials of our class $\cG$. The procedure can then be iterated by replacing $H$ by $e^{W}H e^{-W}$. The convergence of the transformed $H$ to a block diagonal operator which, because of the degeneracy, commutes with $H_{La}$  is quadratic.

We remark that in order to be really applicable to the quantum Hall effect our result should be extended to a more  general class of  potentials than stated above. This should  in principle be possible. However,  a delicate control of $\lbrack H_{La},V\rbrack$ is needed. The method does not work for the extensively studied purely periodic problem $V({q_{1}},{q_{2}})=\cos{q_{1}}+\cos{q_{2}}$.

The plan of the paper is to set up the iterative algorithm in section 2. The class of potentials $\cG$ and control the necessary norms will be discussed in section 3. Section 4 contains the proof of  theorem \ref{thm:main} and a discussion related special cases.

\section{An iterative partial diagonalization algorithm}\label{sec:properties}
As discussed in the introduction, the task is to partially diagonalize the operator $H_{La}+V$. 

Recall that $H_{La}=\sum_{n\in\bN_{0}}\left(n+1/2\right)P^{La}_{n}$ with infinite dimensional projections $P^{La}_{n}$. We consider  $H$ which is of the same type as $H_{La}$; in  order to cover situations where $H$ is already an effective Hamiltonian at finite order we assume that it has either a finite   number of bands $\sigma_{n}$ or an infinite number such that
\[dist\left(\sigma_{n}, \sigma_{m}\right)\ge \fg\vert n-m\vert:\]

\begin{defi}
A selfadjoint operator $H$ is of class $\cC_{\fg}$ for a $\fg>0$ if for a complete family of orthogonal, mutually disjoint projections $\{P_{n}\}_{n\in I \subset\bN_{0}}$ which commute with $H$, it holds for 
\[\sigma_{n}:=spect\left(P_{n}HP_{n}\rstr_{Ran P_{n}}\right):\]
\[\min\sigma_{n+1}-\max\sigma_{n}\ge \fg.\]
\end{defi}

\begin{figure}[hbt]
\centerline {
\includegraphics[height=1.2cm]{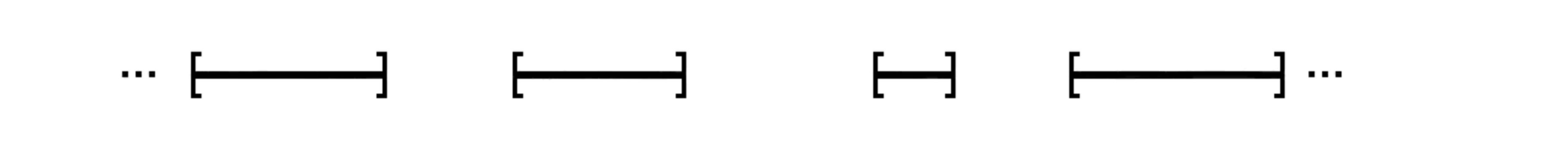}
}
\caption{Typical spectrum of $H$}
\label{fig:spectrum}
\end{figure}

For the same family of projections $P_{n}$ and a bounded operator $V$ we use the notation 
\[\mD V:=\sum_{n\in I}P_{n}VP_{n}, \qquad \mO V:=V-\mD V.\]
\\
To organize the estimates we shall make frequently use of the notations :
\[\langle a\rangle :=\max(1,\vert a\vert), \qquad \Vert\ \Vert\textrm{ : the operator norm, and}\]
\[\Vert A \Vert_l:=\sup_{n,m\in I} \langle n-m\rangle ^l \Vert P_n A P_m\Vert.\]


We prove (extending \cite{dlsv})
\begin{theorem}\label{thm:algo}
Let $H\in\cC_{\fg}$ and $V$ be a bounded selfadjoint operator such that 
\[ \Vert V\Vert_{1}\le\frac{\fg}{8}.\]
Then there exists a unitary $\cU$ such that $\cU^{-1} D(H)\subset D(H)$ with the property  that for 
\[H_{\infty}:=\cU\left(H+V\right)\cU^{-1}, \qquad D(H_{\infty})=D(H)\]
it holds
\[\lbrack H_{\infty},P_{n}\rbrack=0.\]
\end{theorem}

{\bf Proof.} 
Define $H_0 := H+V =\mD H_0 +\mO H_0$. Assume $\Vert V\Vert_{1}\le\frac{\fg}{8}$. Then
$\mD H_0\in \cC_{1/4}$, $\Vert \mO H_0\Vert_1 =\Vert OV \Vert_1 <\infty$. By lemma \ref{lemma:gamma} there exists a bounded solution $W_0$ of
\[\lbrack \mD H_0,W_0\rbrack=\mO H_0, \qquad \mD W_0=0.\]
Define $\cU_0:= e^{W_0}$.\\
Remark : $\cU_0$ is unitary, $D(H_0)\subset D(H),\, W_0 D(H_0)\subset D(H_0),\, \cU_0D(H_0)\subset D(H_0)$.\\
Suppose that for  $s\in\bN$ diagonalization has been done up to $H_s,\, \cU_{s-1}$ such that $\Vert\mD H_s-H\Vert\le\fg/4$, $\Vert \mO H_s\Vert_1<\infty$. 
To go to step $s+1$, use lemma \ref{lemma:gamma} to solve
\[\lbrack \mD H_s,W_s\rbrack=\mO H_s, \qquad \mD W_s=0\]
for a bounded $W_s$ and define $\cU_s:=e^{W_s}\cU_{s-1}$, 
$$H_{s+1}:= e^{W_s} H_s e^{-W_s}=e^{L_{W_s}}(H_s)=\sum_{k=0}^\infty \frac{L_{W_s}^k(H_s)}{k!} $$
with the notation $L_W(A):= [W,A].$
Now
$$L_{W_s}(\mD H_s)=-\mO H_s$$
thus for $k\ge1$
$$L_{W_s}^k(H_s)=-L_{W_s}^{k-1} (\mO H_s)+L_{W_s}^k (\mO H_s)$$
so
\begin{equation}\label{eq:phi}
H_{s+1}=\mD H_s+\phi(L_{W_s})(\mO H_s)
\end{equation}
with
$$\phi(x):=e^x-\frac{1}{x} (e^x-1)=\sum_{k=1}^\infty \frac{k}{(k+1)!}x^k \qquad(x\in\bR).$$
It holds by corollary \ref{coro:commutatornorm}
$$\Vert L_{W_s}(\mO H_s)\Vert_1 \le \Vert W_s \mO H_s\Vert_1+\Vert  \mO H_s W_s\Vert_1\le 2 K\Vert W_s\Vert_2 \Vert \mO H_s\Vert_1.$$
Thus by induction and lemma \ref{lemma:gamma}
$$\Vert L_{W_s}^k(\mO H_s)\Vert_1\le  \left( \frac{\pi 2 K}{ \fg}\right)^k\Vert\mO H_s\Vert_1^{k+1}$$
which implies 
\begin{eqnarray*}
\Vert\phi(L_{W_s})(\mO H_s)\Vert_1& \le&\sum_{k=1}^\infty \frac{k}{(k+1)!}\left(\frac{2\pi K}{\fg}\Vert \mO H_s\Vert_1\right)^k\Vert \mO H_s\Vert_1\\
&=&\frac{\fg}{2\pi K}\psi(\frac{2\pi K}{\fg}\Vert \mO H_s\Vert_1)\end{eqnarray*}
with $\psi(x):=x\phi(x)=(x-1)e^x+1\qquad(x\ge0)$. Remark that $\psi$ is nonnegative, $\psi(x)\le x\quad(x<1)$, and that $0$ is a superattractive fixed point. By (\ref{eq:phi})
\[\frac{2\pi K}{\fg}\Vert \mO H_{s+1}\Vert_{1}\le\psi\left(\frac{2\pi K}{\fg}\Vert \mO H_s\Vert_{1}\right)\]
thus 
\begin{equation}\label{eq:os}
\Vert \mO H_s\Vert_{1}\le\lambda^{2^{s}}
\end{equation}
 with $\lambda$ proportional to $\Vert\mO V\Vert_{1}$ small enough. For the diagonal part it holds
\[\Vert\mD H_{s+1}-\mD H_{s}\Vert=\Vert\mD H_{s+1}-\mD H_{s}\Vert_{1}\le\Vert\phi\left(L_{W_{s}}\right)(\mO H_{s})\Vert_{1}\le\frac{\fg}{2\pi K}\psi(\frac{2\pi K}{\fg}\Vert \mO H_s\Vert_1)\]
and thus for an $x$ proportional to $\Vert\mO V\Vert_{1}$
\[\Vert\mD H_{s+1}-H\Vert\le\frac{\fg}{8}+\frac{\fg}{2\pi K}\int_{0}^{x}\psi\le\frac{\fg}{4}\]
so the iteration is well defined. 

In particular $s\mapsto\mD H_{s}-H$ is a norm convergent sequence. (\ref{eq:os}) and Lemma \ref{lemma:gamma} imply that $\Vert W_{s}\Vert$ converges superexponentially to $0$. By (\ref{eq:phi}) this implies in turn as $\mO V$ is bounded and $\Vert\mO V\Vert_{1}$ is small enough:
\[\Vert\mO H_{s+1}\Vert\le\Vert\phi\left(L_{W_{s}}\right)(\mO H_{s})\Vert\le\underbrace{\sum\frac{k}{(k+1)!}2^{k}\Vert W_{s}\Vert^{k}}_{\le 1}\Vert\mO H_{s}\Vert\]
thus $\mO H_{s}$ converges in operator norm and furthermore as\\ $\Vert\cU_{s+1}-\cU_{s}\Vert=\Vert e^{W_{s+1}}-\bI\Vert$ one concludes that $\cU_{s}\to_{\Vert .\Vert}\cU$. By construction $H_{\infty}=\mD H_{\infty}$ which commutes with the projections. 
\ep

We now  prove some claims which where used in the preceding proof: 
the existence of a  bounded solution of the commutator equation is assured by

\begin{lemma}\label{lemma:gamma}
Let $H\in\cC_{\fg}$ and $V$ a bounded selfadjoint operator such that $P_n V P_n=0\,\, \forall n$ and such that $\Vert V\Vert_1<\infty$. Then there exists a bounded antiselfadjoint $W$ such that
\[\lbrack H,W\rbrack=V, \qquad \mD W=0\]
such that 
\[\Vert W\Vert\le\frac{\pi\zeta(2)}{\fg} \Vert V\Vert_1.\]
\[\Vert W\Vert_2\le\frac{\pi}{2\fg} \Vert V\Vert_1.\]
\end{lemma}

{\bf Proof.}
By \cite{br} for bounded operators $A,B,C$ there exists a solution $X$ of $AX-XB=C$ such that 
\[\Vert X\Vert\le\frac{\pi}{2}\frac{\Vert C\Vert}{dist\left(spect(A),spect(B)\right)}.\]
It follows that there exists $W_{nm}$ such that
\[P_{n}HP_{n}W_{nm}-W_{nm}P_{m}HP_{m}=P_{n}VP_{m}\]
with $W_{nm}=P_{n}W_{nm}P_{m}$  and
\[\Vert W_{nm}\Vert\le\frac{\pi}{2}\Vert P_{n}VP_{m}\Vert \frac{1}{\fg\langle n-m\rangle}\le\frac{\pi}{2\fg}\frac{\Vert V\Vert_1}{\langle n-m\rangle^{2}}.\]
Now define $W:=\sum_{n\neq m}W_{nm}$ in norm convergence, then
\[\Vert W\Vert\le \sup_{n}\sum_{m}\Vert P_{n}WP_{m}\Vert=\sup_{n}\sum_{m}\frac{1}{\langle n-m\rangle^{2}}\frac{\pi\Vert V\Vert_1}{2\fg}\]
from which the claim follows.\ep

\begin{lemma}\label{lem:convolution}
Let $n,m\in \bN_0$ and $K:=3+2\zeta(2)$. It holds: 
\begin{equation}\label{eq:convolution}
\sum_{j\ge0, j\neq n,j\neq m} \frac{1}{\langle j-n\rangle^2\langle j-m\rangle} \le \frac{K}{\langle m-n\rangle}.
\end{equation}
\end{lemma}

{\bf Proof} : The case where $n=m$ is evident. In what follows, we will simply write $j$ as the index for the sum except of $j\ge0, j\neq n,j\neq m$.
\begin{eqnarray*}
\sum_{j} \frac{1}{\langle j-n\rangle^2\langle j-m\rangle} &=&\frac{1}{\langle m-n\rangle} \left (\sum_{j} \frac{\langle j-n+m-j\rangle}{\langle j-n\rangle^2\langle j-m\rangle}\right ) \\
&\le&\frac{1}{\langle m-n\rangle} \bigg ( \sum_{j}\frac{1}{\langle j-n\rangle^2}+\sum_{j}\frac{1}{\langle j-n\rangle\langle j-m\rangle}\bigg ).
\end{eqnarray*}
The left term of the r.h.s. is bounded by $2 \zeta(2)$. It remains to prove that the second term of the r.h.s. is bounded. Recall the  series expansion of  the digamma function $\psi_0$  :
\begin{equation}\label{eq:formuledigamma} \psi_0(x+1)+\gamma=\sum_{j=1}^\infty \frac{x}{j(j+x)}\qquad  (x\in \bN).\end{equation}
Now, since it is symmetric in $m$ and $n$, we can assume  that $m>n$ and  define $a:=m-n>1$. Then the second term of the r.h.s. becomes :
\begin{eqnarray*}
\sum_{j}\frac{1}{\langle j-n\rangle\langle j-m\rangle}&=& \sum_{j\ge -n}\frac{1}{\langle j \rangle \langle j-a\rangle}\\
&=&\sum_{j=-n}^{a-1}\frac{1}{\langle j \rangle \langle j-a\rangle}+\sum_{j\ge a+1} \frac{1}{\langle j \rangle \langle j-a\rangle}\\
&=&\sum_{j=-n}^{a-1}\frac{1}{\langle j \rangle\langle j-a\rangle}+\frac{\gamma+\psi_0(1+a)}{a}
\end{eqnarray*}
where we used (\ref{eq:formuledigamma}). But
\begin{eqnarray*}
\sum_{j=-n}^{a-1}\frac{1}{\langle j \rangle \langle j-a\rangle}&=&\sum_{j=1}^{n}\frac{1}{\langle j \rangle \langle j+a\rangle}+\sum_{j=1}^{a-1}\frac{1}{\langle j \rangle \langle j-a\rangle}\\
&\le&\frac{\gamma+\psi_0(1+a)}{a} +\vert a-1 \vert \sup_{j\in [1,a-1]\cap \bN} \frac{1}{\vert j \vert \vert j-a\vert}.
\end{eqnarray*}
Since $a>1$, both of these terms are bounded by 1 so the claim of the lemma follows.\ep

We have the following corollary :
\begin{corollary}\label{coro:commutatornorm}
For operators $A$ and $B$ such that $\Vert A \Vert_2<\infty$ and $\Vert B\Vert_1<\infty$ it holds
$$\Vert AB \Vert_1 \le K \Vert A\Vert_2\Vert B\Vert_1$$
where $K$ was defined in \ref{lem:convolution}.
\end{corollary}

{\bf Proof} : Since $\lbrace P_n\rbrace_{n\in I}$ is a complete family of orthogonal and mutually disjoint projectors, we can write
$$P_n AB P_m= \sum_{l\in I} (P_n\, A\, P_l)(P_l\, B\, P_m).$$
Thus
$$\Vert AB \Vert_1 \le \sup_{n, m}\, \langle m-n\rangle \sum_{l\ge 0} \frac{\Vert A \Vert_2}{\langle n-l\rangle^2} \frac{\Vert B\Vert_1}{\langle l-m\rangle} $$
The result follows from lemma \ref{lem:convolution}.\ep

\section{The class $\cG$}
We show that the basic decay estimate is satisfied for potentials in the class $\cG$ defined in the introduction and give some examples.
\begin{propo}\label{propo:cG}For $V\in\cG$ and $P_{n}$ the eigenprojections of $H_{La}$ on its n-th level it holds in operator norm on $L^{2}(\bR^{2})$ for a $d>0$ and all $n,m\in\bN_{0}$:
\[\Vert P_{n}VP_{m}\Vert\le\frac{d}{\langle n-m\rangle}.\]
\end{propo}

{\bf Proof.}
\begin{eqnarray*}
\Vert  P_n V P_m \Vert &=& \Vert P_n \int g(.-y) d\mu(y) P_m\Vert\\
&\le& \int \vert d\mu(y) \vert \,\Vert P_n g(.-y)P_m\Vert.
\end{eqnarray*}
Consider the unitary magnetic translations on $L^2(\bR^2)$ defined for a $a\in\bR^2$ by
$$T(a) \psi(q)=e^{\frac{i}{2} q\wedge a} \psi(q-a).$$
It holds $[ T(a),H_{La}]=0$ and $T(a)\, g \,T^*(a) = g(.-a)$. Thus
$$\Vert P_n\, g(.-y) P_m\Vert = \Vert P_n\, g \,P_m\Vert\quad \forall y\in\bR^2$$
and the result follows from Proposition \ref{propo:gaussian} to be proven below. \ep

Remark that $V=g\ast\mu$  extends necessarily to an entire analytic function; its Fourier Transform $\widehat{V}$ has the property that $\widehat{V}(p)\exp{\frac{p^{2}}{2}}$ is the Fourier Transform of a finite measure. We elaborate on this in order to point out that $\cG$ contains sufficiently many potentials to be of interest for applications to the Quantum Hall effect. 

\begin{defi}
Denote by $\cA$ the class of functions $V:\bR^2 \to \bR$ such that
\begin{enumerate}
\item V has an extension to an entire function on $\bC^2$,
\item$\bR^2 \ni y \mapsto e^{-\frac{y^2}{2}} V(i y)\in L^1(\bR^2)$,
\item for $\widetilde{V}(p):= e^{\frac{p^2}{2}} \int_{\bR^{2}} e^{-i p y} e^{-\frac{y^2}{2}} V(i y) \frac{dy}{(2\pi)^2}$ it holds: $\widetilde{V}\in L^1(\bR^2)$.
\end{enumerate}
\end{defi}

\begin{propo}For $V\in\cA$ it holds
$$V=g\ast (\widetilde{V} dq).$$
\end{propo}

{\bf Proof.}
By Fourier's theorem, it holds for $q\in\bR^2$
$$e^{-\frac{q^2}{2}} V(i q)= \int _{\bR^2\times\bR^{2}} e^{i p (q-y)} e^{-\frac{y^2}{2}} V(i y) \frac{dy}{(2\pi)^2} dp.$$
Thus
$$V(iq)=\int_{\bR^2} e^{\frac{(q+ip)^2}{2}} \widetilde{V}(p) dp$$
and the claims follows as both sides are analytic in $q$.\ep

\bigskip

We list some examples of potentials in $\cG$ which contain in particular Anderson type models on a finite portion of the probe.

\begin{corollary} $\cG$ contains
\begin{enumerate}
\item for $p$ a real polynomial, $\alpha \in (0,1)$, $k_1,k_2\in \bR$:
 \[p(q_1,q_2, e^{ik_1 q_1},e^{i k_2 q_2}) e^{-\alpha \frac{q^2}{2}},\] 
 \item $\sum_{i\in\bZ^2} \mu_i \, g(q-i)$ with $\mu\in l^1(\bZ^2,\bR)$.
\end{enumerate}
The function $q\mapsto e^{i k q}$ for $k\in \bR^2$ does not belong to $\cG$.
\end{corollary}

{\bf Proof.}
For $1.$ it is sufficient to verify that 
$$\bR \ni y \mapsto e^{-\frac{y^2}{2}} f(i y)\in L^1(\bR)$$ and $$\bR \ni x\mapsto e^{\frac{x^2}{2}} \int e^{-i x y}e^{-\frac{y^2}{2}} f(i y)dy\in L^1(\bR)$$
for $f(y)=y^n e^{-\alpha \frac{y^2}{2}}$ and $f(y) = e^{i k y}e^{-\alpha \frac{y^2}{2}}$, $n\in\bN$ and $k\in\bR$.\\
This is by standard properties of the Fourier transform. In the first case,
$$\vert f(iy) e^{-\alpha \frac{y^2}{2}}\vert = \vert y\vert^n e^{-(1-\alpha) \frac{y^2}{2}} \in L^1(\bR)$$
and
$$\vert e^{\frac{x^2 }{2}} \int e^{-i x y}y^n e^{-(1-\alpha)\frac{y^2}{2}} dy\vert=\frac{1}{\sqrt{1-\alpha}}\vert poly(x) e^{\frac{x^2 }{2}} e^{-\frac{1}{1-\alpha}\frac{x^2}{2}}\vert\in L^1(\bR).$$
In the second case
$$\vert f(iy) e^{-\alpha \frac{y^2}{2}}\vert \le e^{\vert k\vert \vert y\vert}  e^{-(1-\alpha) \frac{y^2}{2}}\in L^1(\bR)$$
$$\vert e^{\frac{x^2 }{2}} \int e^{-i x y} e^{- k y} e^{-(1-\alpha)\frac{y^2}{2}} dy\vert =\frac{1}{\sqrt{1-\alpha}}e^{\frac{k^2}{2(1-\alpha)}} e^{\frac{x^2}{2}} e^{-\frac{1}{1-\alpha}\frac{x^2}{2}}\in L^1(\bR)$$
In $2.$, one deals with the pure point measure
$$\sum_{i\in\bZ^2} \mu_i \delta(x-i).$$
Finally, one has
$$e^{ikx}=g\ast \mu$$
with $\mu=e^{-ikx} e^{\frac{k^2}{2}}dx$ which is not a finite measure.\ep

It remains to treat the case of the gaussian potential $g$ which turns out to be non trivial. We follow the strategy designed in \cite{w08}.
\begin{propo}\label{propo:gaussian}For $g(q)=\exp{\left(-\frac{q^{2}}{2}\right)},\quad (q\in\bR^{2})$ and $P_{n}$ the eigenprojections of $H_{La}$ on its n-th level it holds in operator norm on $L^{2}(\bR^{2})$ for a $c>0$ and all $n,m\in\bN_{0}$:
\[\Vert P_{n}gP_{m}\Vert\le\frac{c}{\langle n-m\rangle}.\]
\end{propo}
{\bf Proof.} We use the representation of $P_{n}$ by angular momentum eigenfunctions
\[P_{n}=\sum_{l\ge -n}|\psi_{n,l}\ket\bra\psi_{n, l}|\]
\begin{equation}\label{def:eigenfunctions}
\psi_{n,l}(r,\Theta):=(-1)^{n}\sqrt{\frac{n!}{2^{l}(n+l)!}}r^{l}e^{i\Theta l} L_{n}^{l}\left(\frac{r^{2}}{2}\right)\frac{e^{-\frac{r^{2}}{4}}}{\sqrt{2\pi}}
\end{equation}
where the Laguerre polynomials are defined by 
\[L_{n}^{l}(x):=\sum_{j=0}^{n}\frac{(-x)^{j}}{j!}
\left(\begin{array}{c}n+l \\n-j   \end{array}\right)  \quad (l\ge0)\]
\[
L_{n}^{l}(x):=\frac{(n+l)!}{n!}(-x)^{\vert l\vert}L_{n+l}^{\vert l\vert}(x)\quad (0\ge l\ge-n).
\]
Then 
\[P_{n}gP_{m}=\sum_{l\ge-n\land m}|\psi_{n,l}\ket\bra\psi_{n,l}, g\psi_{m,l}\ket\bra\psi_{m,l}|\]
thus
\[\left\vert\bra\psi, P_{n}g P_{m}\varphi\ket\right\vert\le\sup_{l\ge-n\land m}\left\vert \bra\psi_{n,l}, g\psi_{m,l}\ket\right\vert\Vert\psi\Vert\Vert\varphi\Vert\]
and the claim follows from equation (\ref{eq:matrixelements}) and estimate (\ref{eq:majorationelementmatrice}) to be proven in the following two lemmas.
\ep
\begin{lemma}
For $g(q)=\exp{\left(-\frac{q^{2}}{2}\right)},\quad (q\in\bR^{2})$ and $\psi_{n,l}$ defined in (\ref{def:eigenfunctions}) it holds for $n, m\in\bN_{0}, l\ge -(n\land m)$
\begin{equation}
\label{eq:matrixelements}\left\vert\bra\psi_{n,l}, g\psi_{m,l}\ket\right\vert=\frac{1}{2^{l+m+n+1}}\frac{(l+m+n)!}{\sqrt{(l+m)!(l+n)!n!m!}}.
\end{equation}

\end{lemma}
{\bf Proof.} By definition
\[\left\vert\bra\psi_{n,l}, g\psi_{m,l}\ket\right\vert=\frac{1}{2^{l}}\sqrt{\frac{n!m!}{(l+n)!(l+m)!}}\int_{0}^{\infty}e^{-r^{2}}r^{2l}L_{n}^{l}L_{m}^{l}\left(\frac{r^{2}}{2}\right)r dr.\]
Consider first $l\ge0$. To study the dependence of the integral on $l, m, n$ we use that the family $n\mapsto L_{n}^{l}(x)$ is orthogonal in $L^{2}\left(\bR_{+}, d\nu_{l}\right)$, $d\nu_{l}:=x^{l}e^{-x}dx$ and the identity: 
\[L_{n}^{l}L_{m}^{l}\left(\frac{x}{2}\right)=\sum_{s\ge0}B_{s}^{n, m, l}L_{s}^{l}(x).\]
As $L_0^l\equiv 1\, \forall l$, one has
$$\int_0^\infty L_n^l L_m^l(\frac{x}{2})d\nu_l(x) = \sum_{s\ge0} B_s^{n,m,l} \int_0^\infty L_s^l L_0^l(x)d\nu_l(x)$$
$$ = B_0^{n,m,l}\int_0^\infty d\nu_l(x)= B_0^{n,m,l}\Gamma(l+1)$$
It was proven in \cite{ca} that
\begin{eqnarray*}
g_s(x,y,l)&=&\dfrac{\left( \frac{x}{2(1-x)} +\frac{y}{2(1-y)}\right)^s}{(1-x)^{l+1}(1-y)^{l+1}\left(1+\frac{x}{2(1-x)}+\frac{y}{2(1-y)}\right)^{l+s+1}}\\
&=&\sum_{n,m} B_s^{n,m,l}x^m y^n.
\end{eqnarray*}
Thus
$$B_0^{n,m,l}=\frac{1}{n!m!} \partial_x^m \partial_y^n g_0(x,y,l)\bigg\vert _{x=y=0} = \frac{1}{2^{m+n}} \frac{(l+m+n)!}{l!\,m!\,n!}$$
from which the claim follows for $l\ge0$.\\
Now for $l<0$, one has
$$\psi_{n,l} =\overline{\psi}_{n+l,-l}$$
thus
$$ \vert \langle \psi_{n,l}, V \psi_{m,l}\rangle\vert = \vert \langle \psi_{n+l,-l}, V\psi_{m+l,-l}\rangle\vert $$
and the result follows for $l < 0$.\ep

The following lemma might be known to probabilists, we know of  no reference though.
\begin{lemma}
For a $c>0$, it holds
\begin{equation}
\label{eq:majorationexponentielle}\frac{(m+n)!}{2^{m+n} n!m!}\le c \frac{m+n}{\sqrt{(m+n)^2-(m-n)^2}} e^{-\frac{m-n}{2(m+n)}}\quad (\bN\ni m,n\ge1),
\end{equation}
\begin{equation}
\label{eq:majorationm-n}\frac{(m+n)!}{2^{m+n} n!m!}\le\frac{c}{\langle m-n\rangle}\hfill\qquad (m,n\in \bN_0),
\end{equation}
\begin{equation}
\label{eq:majorationelementmatrice}\frac{(l+m+n)!}{2^{l+m+n+1}\sqrt{(l+m)!(l+n)!n!m!}}\le\frac{c}{\langle m-n\rangle}\qquad(m,n\in \bN_0,l\ge -m\wedge n).
\end{equation}
\end{lemma}
{\bf Proof.} Denote by $C$ a positive constant whose value may change from line to line.\\
We use Stirling's and a concavity inequality :
\begin{equation}\label{eq:stirling}\frac{1}{C}\le \frac{n!}{\left( \frac{n}{e}\right)^n \sqrt{n}}\le C \hspace{2cm} (n\ge1)\end{equation}
\begin{equation}\label{eq:logcarre}(1+x)\log(1+x)+(1-x)\log(1-x)\ge x^2\hspace{1cm} (x\in [0,1])\end{equation}
to estimate
$$a_{m,n}:= \frac{(m+n)!}{2^{m+n} n!m!}.$$
For $m,n\ge 1$ it holds by (\ref{eq:stirling}) :
$$a_{m,n}\le C \frac{(m+n)^{(m+n)}}{m^m\,n^n\,2^{m+n}}\sqrt{\frac{1}{n}+\frac{1}{m}}\le C \frac{a^a}{(a+b)^{\frac{a+b}{2}}(a-b)^{\frac{a-b}{2}}} \sqrt{\frac{a}{a^2-b^2}}$$
with $a:=m+n$, $b:=m-n$ in
$$G:=\left\lbrace (a,b)\in \bZ^2; a \ge2, \vert b \vert \le \vert a-2\vert \right\rbrace.$$
Note that the case $n=m$ follows from the first inequality of the previous line.
Using (\ref{eq:logcarre})   with $x:=\frac{b}{a}$ it follows
$$a_{m,n} \le \sqrt{\frac{a}{a^2-b^2} }\,e^{-\frac{b^2}{2a}}$$
which implies (\ref{eq:majorationexponentielle}). Consider now
$$ G_<:= G\,\cap \left\lbrace (a,b)\in \bZ^2; \frac{b^2}{a^2}<\frac{1}{2}\right\rbrace,$$
then
$$\langle m-n\rangle ^2\, a_{m,n}^2\le C \frac{1}{1-\frac{b^2}{a^2}}\frac{b^2}{a} e^{-\frac{b^2}{2a}}<2\hspace{1cm}((a,b)\in G_<).$$
In $G\setminus G_<$, it holds
$$\frac{a}{2}\le \frac{b^2}{a}\le a\quad \textrm{as well as}\quad 1-\frac{b^2}{a^2}\ge \frac{2}{a}$$
thus
$$\langle m-n\rangle ^2 \,a_{m,n}^2 \le C \,a^2 \,e^{-\frac{a}{2}}\le C$$
which proves (\ref{eq:majorationm-n}) as it is evident for $n=0$.\\
Denote now
$$b_{l,m,n} :=\left( \frac{(l+m+n)!}{2^{l+m+n}}\right)^2 \frac{1}{(l+m)!(l+n)!m!n!}.$$
From the identity $b_{l,m,n}=a_{l+m,n}\, a_{l+n,m}$ it follows for $l+m,l+n,m,n\ge1$
$$b_{l,m,n}\le C \sqrt{\frac{a}{a^2-b^2}} \sqrt{\frac{a}{a^2-c^2}} e^{-\frac{b^2+c^2}{2a}}$$
with $a:= l+m+n, b:=l+m-n, c:=l+n-m, (a,b)\in G\textrm{ and }(a,c)\in G$. It follows
$$\langle m-n\rangle ^2 b_{l,m,n}\le C \bigg\vert \frac{b-c}{\sqrt{a}}\bigg\vert^2\frac{1}{\sqrt{1-\frac{b^2}{a^2}}\sqrt{1-\frac{c^2}{a^2}}}e^{-\frac{b^2+c^2}{2a}}.$$
In $G$, one has $\frac{1}{1-\frac{b^2}{a^2}}\le \frac{a}{2}$ so if $\frac{b^2}{a^2}\ge \frac{1}{2}$ then $\frac{a}{2}\le \frac{b^2}{a}\le a$ and, denoting any polynomial by $``poly"$ :
$$\langle m-n\rangle  ^2 b_{l,m,n} \le C \,poly(\frac{c}{\sqrt{a}},\sqrt{a}) e^{-\frac{a}{2}}e^{-\frac{c^2}{2a}}\le C.$$
Now if $\frac{b^2}{a^2}< \frac{1}{2}$ then either $\frac{c^2}{a^2}\ge \frac{1}{2}$ and
$$\langle m-n\rangle  ^2 b_{l,m,n} \le C poly( \sqrt{a}, \frac{b}{\sqrt{a}}) e^{-\frac{b^2+c^2}{2a}}\le C$$
or $\frac{c^2}{a^2}< \frac{1}{2}$ and
$$\langle m-n\rangle  ^2 b_{l,m,n} \le C poly(\frac{b}{\sqrt{a}}, \frac{c}{\sqrt{a}}) e^{-\frac{b^2+c^2}{2a}}\le C.$$
For the cases where one of $l+m, l+n,m,n$ is zero, remark first that
$$b_{l,n,n}=\vert a_{l+n,n}\vert^2\le \frac{C}{\langle l\rangle^2}\le C$$
by (\ref{eq:majorationm-n}) and secondly that for $l+m=0, n\ge m$
$$\langle m-n\rangle  ^2 b_{l,m,n} = \frac{1}{2^n}\langle m-n\rangle  ^2  a_{l+n,m}\le \frac{\langle m-n\rangle  ^2 }{2^{n-m}} \frac{C}{2^m \langle n-2m\rangle}\le C$$
which covers all cases. \ep
\section{Application of the algorithm}
We proof Theorem \ref{thm:main} then give an illustration.

{\bf Proof.} (of Theorem \ref{thm:main})

Choose $H=H_{La}$ and $\lbrace P_{n}\rbrace_{n\in\bN_{0}}$  its eigenprojections. Then $H_{La}\in\cC_{1}$. By proposition \ref{propo:cG} $\Vert V\Vert_{1}$ is finite for $V\in\cG$. So for a $V\in\cG$ with $\Vert V\Vert_{1}\le\frac{1}{8}$ by theorem \ref{thm:algo} there exists $\cU$ unitary such that $\lbrack\cU\left(H_{La}+V\right)\cU^{-1},P_{n}\rbrack=0$ thus $\lbrack\cU\left(H_{La}+V\right)\cU^{-1},H_{La}\rbrack=0$.\ep

\subsection{Quadratic Hamiltonians}

We discuss the case where $V$ is a polynomial of degree at most 2 for sufficiently high magnetic field. Though this case is not covered directly by Theorem \ref{thm:main}, the iterative algorithm can be applied to the hamiltonian matrix which defines the operator. This results in the construction of an integral of motion which is the quantization of a classical integral, independent of the hamiltonian function.
The following operations are to be understood first on vectors in $\cS(\bR^2)$ then on the appropriate extensions.
Denote $D:=-i\nabla$, the velocity and center operators
$$ v:= D-\frac{q^\perp}{2},\quad c:=-D^\perp+\frac{q}{2}$$
and recall the commutation relations
$$[v_1,v_2]=i\,\, ,\quad[c_2,c_1]=i\,\, ,\quad[c_i,v_j]=0.$$
The linear case is trivial, nevertheless  it is instructive :
$$ V(q)=-\langle E,q\rangle=-(E_1q_1+E_2q_2),$$
define $W_0:=i\langle E,v\rangle,\, \cU_0=e^{W_0}$. From the Weyl relations
$$e^{i\langle E,v\rangle }\,v\, e^{-i\langle E,v\rangle}=v-E^\perp$$
it follows 
\begin{propo}
For $E\in\bR^2$, $V(q)=-\langle E,q\rangle$, $\cU_0=e^{i\langle E,v\rangle }$, it holds

\begin{enumerate}
\item $\cU_0(H_{La}+V)\cU_0^{-1}=H_{La}-\langle E,c\rangle -\frac{1}{2}E^2=\mD{H}-\frac{1}{2}E^2$
\item $\cU_0^{-1}\,H_{La}\,\cU_0=\frac{1}{2} (v+E^\perp)^2.$
\end{enumerate}
\end{propo}
Now consider the quadratic case, $V(q)=\frac{1}{2}\langle q,V''q\rangle$
for a real symmetric $2\times2$ matrix $V''$.

The Hamiltonian is

\begin{eqnarray*}
H&=&\frac{1}{2}\left(D-\frac{q^\perp}{2}\right)^2+\frac{1}{2}\left\langle q,V''q\right\rangle\\
&=&\frac{1}{2}\left\langle \begin{pmatrix} q\\D\end{pmatrix},\begin{pmatrix}0&\bI\\-\bI&0\end{pmatrix}\bH\begin{pmatrix} q\\D\end{pmatrix}\right\rangle\rangle
\end{eqnarray*}
with
$$\bH=\begin{pmatrix}{\sigma^t/2}&-\bI\\ \bI/4+V''&{\sigma^t/2}\end{pmatrix}$$

where we denote $\sigma=\begin{pmatrix}0 & 1 \\ -1 & 0\end{pmatrix}$ and $\bI$ the $2\times2$ identity matrix.

$\bH$ is a hamiltonian matrix with respect to the symplectic structure defined by $\begin{pmatrix}0&\bI\\-\bI&0\end{pmatrix}$. Its eigenvalues are $$\lbrace\pm\underbrace{i\sqrt{P+Q}}_{:=\lambda},\pm\underbrace{\sqrt{Q-P}}_{:=\mu}\rbrace$$
with $P:=1+tr V'',\, Q=P^2-4\det V''$.
For $V''$ small enough, $P>0, Q>0$ thus $\lambda\in i\bR$.
In the case of an hyperbolic fixed point, $\det V''<0$ so $\mu\in\bR$. In the elliptic case, $\mu\in\bR$ if $\det V''$ is small enough and in the parabolic case, $\mu=0$. In all three cases one knows from  normal form theory  (see \cite{wi}) that there exists a symplectic transformation decoupling the degrees of freedom. We state the explicit result for the cases of the quantum dot and antidot, i.e.: $V''=\pm \varepsilon^{2}\bI$ which one verifies by direct calculation:

\begin{propo}
Let $1/2>\varepsilon>0$, $V(q)=\pm\frac{\varepsilon^{2}}{2}\left(q_{1}^{2}+q_{2}^{2}\right)$. Define $\Omega:=\sqrt{1\pm4\varepsilon^{2}}$ and the unitary  $\cU\psi(q)=\frac{1}{\sqrt{\Omega}}\psi(q/\sqrt{\Omega})$, then  it holds

\begin{enumerate}
\item $\cU(H_{La}+V)\cU^{-1}=\frac{1+\Omega}{2}H_{La}+\frac{\Omega-1}{2}\frac{c^{2}}{2}=\mD H+\frac{\Omega-1}{2}(H_{La}+\frac{c^{2}}{2})\mp\frac{\varepsilon^{2}}{2}\frac{c^{2}}{2},$

\item and for the constant of motion:
\[\cU^{-1}\,H_{La}\,\cU=\frac{1}{2}\left(\frac{1}{\sqrt{\Omega}}D-\sqrt{\Omega}\frac{q^\perp}{2}\right)^2.\]
\end{enumerate}
\end{propo}


\begin{thebibliography}{xxxxxxx}
 

 \bibitem[ABJ]{abj}Asch, J., Bourget, O., Joye, A.: Localization Properties of the Chalker-Coddington Model.  {\it http://arxiv.org/abs/1001.3625v1} (2010).
 
\bibitem[ADE]{ade} Asch, J., Duclos, P., Exner, P.:Stability of driven systems with growing gaps, quantum rings, and Wannier ladders. {\it J. Statist. Phys.} {\bf 92} (1998), no. 5-6, 1053--1070.
 

\bibitem[ASS]{ass}Avron, J. E., Seiler, R., Simon, B.: Charge deficiency, charge transport and comparison of dimensions.  {\it Comm. Math. Phys.}  {\bf 159} , 399--422,  (1994).

\bibitem[B]{b} Bellissard, J., Stability and instability in quantum mechanics, in Trend and Development of the Eighties (Bielefeld 1982/1983), (S. Albeverio and P. Blanchard, eds), World Scientific, Singapore 1985, pp. 1Ð106.

\bibitem[BDP]{bdp}Br\"uning, J.; Dobrokhotov, S. Yu.; Pankrashkin, K. V. ``The spectral asymptotics of the two-dimensional Schršdinger operator with a strong magnetic field. II''. {\it Russ. J. Math. Phys.} {\bf 9} (2002), no. 4, 400--416.

\bibitem[BG]{bg}Buchendorfer, C., Graf, G.M., ``Scattering of magnetic edge states.''. {\it Ann. Henri PoincarŽ} {\bf 7} (2006), no. 2, 303--333.

\bibitem[BESB]{bsbe}Bellissard, J., van Elst, A., Schulz-Baldes, H.: ``The noncommutative geometry of the quantum Hall effect,'' J. Math. Phys. \textbf{35}, 5373-5451 (1994).

\bibitem[BR]{br} Bhatia,~R., Rosenthal,~P.;
\emph{How and Why to Solve the Operator Equation $AX-XB=Y$}.
Bulletin of the London Mathematical Society 29, 1997. pp. 1-21.

\bibitem[Ca]{ca}Carlitz, L. : ``The product of several Hermite or Laguerre polynomials,'' {\it Monatshefte f\"ur Mathematik} \textbf{66}, 393--396 (1962).

\bibitem[C]{c} Combescure, M.: The quantum stability problem for time-periodic perturbations of the harmonic oscillator,  {\it Ann. Inst. H. Poincare Phys. Theor.} {\bf 47} (1987), 62Ð82; Erratum: {\it Ann. Inst. H. Poincare Phys. Theor.} {\bf 47}(1987), 451Ð454.

\bibitem[CB]{cb} Cary~J.~R, Brizard~A.~J., {}``Hamiltonian
theory of guiding-center motion'', Rev. Mod. Phys. \textbf{81}, 693-738
(2009)

\bibitem[CC]{cc} Chalker, J.T., Coddington, P.D.: Percolation, quantum tunneling and the integer Hall effect, {\it J. Phys. C} {\bf 21}, 2665-2679, (1988).


\bibitem[D\v S]{ds}
P.~Duclos, P.~{\v S\v tov\'\i\v cek}.
\newblock Floquet hamiltonians with pure point spectrum.
\newblock {\em Commun. Math. Phys.}, 177:327--374, 1996.

\bibitem[DL\v SV]{dlsv}
P.~Duclos, O. Lev,  P.~{\v S\v tov\'\i\v cek}, M. Vittot.
 ``Progressive diagonalization and applications'', Proceedings of the Conference ``Operator Algebras \& Mathematical Physics'', Constan\c ta (Roumanie, 2001), R. Purice Ed., Theta Foundation, Bucarest (2003).


\bibitem[G]{g} Graf, G. M.: Aspects of the integer quantum Hall effect. Spectral theory and mathematical physics: a Festschrift in honor of Barry Simon's 60th birthday, 429--442, Proc. Sympos. Pure Math., {\bf 76}, Part 1, Amer. Math. Soc., Providence, RI, (2007)

\bibitem[GWNO]{gwno} Geisel, T., Wagenhuber, J., Niebauer, P., Obermair, G.: Chaotic Dynamics of Ballistic Electrons in Lateral Superlattices and Magnetic Fields, {\it Phys. Rev. Lett.} {\bf 64}, 1531--1534, (1990).

\bibitem[GKS]{gks} Germinet, F., Klein, A., Schenker, J.: Dynamical delocalization in random Landau Hamiltonians, {\it Ann. of Math.} {\bf 166}, 215--244, (2007).

\bibitem[L]{l} Laughlin~R.~B., {}``Quantized Hall conductivity
in two dimensions,'' Phys. Rev.~B \textbf{23}, 5632-5633 (1981)



\bibitem[HS]{hs} Helffer, B., Sj\"ostrand, J.: Analyse semi-classique pour l'\'equation de Harper: (French)  Schršdinger operators(S\o nderborg, 1988), 118--197, Lecture Notes in Phys., 345, Springer, Berlin, 1989. 







\bibitem[Nei]{nei} Neishtadt, A.I.: The separation of motions in systems with rapidly rotating phase, {\it  Journal of Applied Mathematics and Mechanics} {\bf 48}, 133--139, (1984).



\bibitem[N]{n}Nenciu, G. On asymptotic perturbation theory for quantum mechanics: al-
most invariant subspaces and gauge invariant magnetic perturbation theory, {\it J.
Math. Phys.} {\bf 43}, 1273Ð1298, (2002).

\bibitem[PG]{pg} Petschel, G., Geisel, T.: Bloch Electrons in Magnetic Fields: Classical Chaos and Hofstadter's Butterfly, {\it Phys. Rev. Lett.} {\bf 71}, 239--242, (1993).


\bibitem[T]{t} Teufel, Stefan Adiabatic perturbation theory in quantum dynamics. Lecture Notes in Mathematics, 1821. Springer-Verlag, Berlin, 2003. 


\bibitem[Wi]{wi}Williamson, J. On the algebraic problem concerning the normal form of linear dynamical systems. {\it Amer. J. Math.}, {\bf 58} 141Ð163, 1936.

\bibitem[W]{w08} Wang, W.M.: Pure point spectrum of the Floquet Hamiltonian for the quantum harmonic oscillator under quasi-periodic perturbations, {\it Comm. Math. Phys.} {\bf 277, No. 2}, 459 -- 496, (2008).
\end{thebibliography}
\end{document}